\documentstyle[epsfig,preprint,eqsecnum,prb,aps]{revtex}

\begin{document}
\draft
\preprint{NSF-ITP-99-111}


\title{Finite size spectrum, magnon interactions and magnetization   
	of $S=1$ Heisenberg spin chains}
\author{
Jizhong Lou${}^{1}$, 
Shaojin Qin${}^{2,1}$, 
Tai-Kai Ng${}^{3}$, 
Zhaobin Su${}^{1}$,
Ian Affleck${}^{2,4,5}$
}
\address{
${}^{1}$Institute of Theoretical Physics, P. O. Box 2735, 
	Beijing 100080, P. R. China\\
${}^{2}$Department of Physics, University of British Columbia, 
	Vancouver V6T 1Z1, BC, Canada\\
${}^{3}$Department of Physics, 
	Hong Kong University of Science and Technology, Hong Kong\\
${}^{4}$Canadian Institute for Advanced Research, 
	The University of British Columbia, 
	Vancouver, B.C., V6T 1Z1, Canada, \\
${}^{5}$Institute for Theoretical Physics, University of California,
	Santa Barbara, CA93106-4030
}
\date{ \today }
\maketitle

\begin{abstract}
We report on density matrix renormalization-group and analytical work 
on $S=1$ antiferromagnetic Heisenberg spin chains.  We study the finite 
size behavior  within the framework of the non-linear sigma model.  We 
study the effect of magnon-magnon interactions on the finite size 
spectrum and on the magnetization curve close to the critical magnetic 
field, determine the magnon scattering length and compare it to the 
prediction from the non-linear $\sigma$ model.
\end{abstract}


\section{Introduction and Conclusions}
Since Haldane pointed out the difference between integer spin and half 
integer spin Heisenberg chains\cite{hal1}, the richness of the physics 
in the Heisenberg model has attracted much attention.  The $S=1$ 
antiferromagnetic Heisenberg chain is gapped\cite{hal1} and the 
magnitude of the gap has been accurately computed\cite{hus,gol}. With
open boundary conditions (OBC), the Hamiltonian:
\begin{equation}
\label{sham}
 H=\sum_{j=1}^{L-1}{\bf S}_{j}\cdot{\bf S}_{j+1},
\end{equation}
gives low energy effective S=1/2 spins localized near the chain ends 
whose interaction decreases exponentially with system size, resulting 
in an exponentially low lying excited state.
\cite{ken1,hagiwara,aff1,ng0,dit,qin1}.

Many low energy features of the model can be understood using an
approximate mapping onto the non-linear $\sigma$ model (NL$\sigma$M).  
In particular, the low lying excited states are constructed from a 
triplet of interacting bosons (magnons).  There are various types of 
predictions from the NL$\sigma$M for the behavior of S=1 chains that 
have been tested to varying extents numerically or experimentally.  
Some of the predictions are not really specific to the NL$\sigma$M but 
would follow from any relativistic quantum field theory with a triplet 
of massive particles.  These include the relativistic dispersion 
relation\cite{qin2}, exponential temperature ($T$) dependence of the 
specific heat and Bose condensation transition at a critical magnetic 
field\cite{tsv,aff2}.  A qualitative prediction specific to the 
NL$\sigma$M is the absence of boundstates.  Quantitative predictions, 
following from the details of the magnon interactions, have also been 
made concerning the two\cite{mag2} and three\cite{mag3} magnon 
contributions to the neutron scattering cross section.  However, 
perhaps mainly because these contributions are so small, they have, so 
far, not been tested numerically or experimentally.

In this paper we present numerical results on low lying states of the
spin chain with both open and periodic boundary conditions using  the 
density matrix group (DMRG) technique.\cite{wht0}  We keep m=800
states in our DMRG calculations.  The biggest truncation error is
$10^{-8}$ and the largest system size is L=200. These results are 
used to study the interactions of the edge spins with each other and 
with bulk magnons, as well as the interactions of 2 magnons with each 
other.  In particular, we determine numerically one parameter which 
characterizes the low energy magnon interactions, the scattering length 
for two parallel spin magnons.  We deduce how this single parameter 
determines  finite size corrections to the S=2 two magnon state with 
both periodic and open boundary conditions.  We also show that the same 
parameter determines the leading correction to the square root 
singularity of the magnetization at the (lower) critical field and
generalize this results to finite T and length L.  Numerical results on 
all these quantities give a consistent value for the magnon scattering 
length of the S=1 chain, of about -2 lattice constants or about 
$-\xi /3$ where $\xi$ is the correlation length.  On the other hand, 
the exact S-matrix for the NL$\sigma$M gives a scattering length of 
$-2\xi /\pi$, roughly twice as large.  Thus, the NL$\sigma$M does not 
fair well in its first quantitative comparison to the S=1 chain.  This 
is perhaps not terribly surprising given that the mapping only becomes 
exact, at low energies, at large S. Better agreement could be expected
for larger S or for S=1 with a (non-frustrating) ferromagnetic next 
nearest neighbor coupling added to decrease the  bare coupling constant 
in the NL$\sigma$M.  This rather poor agreement suggests that other 
predictions of the NL$\sigma$M, such as the  two\cite{mag2} and three 
magnon\cite{mag3} contributions to the neutron scattering cross section 
may not be very accurate either. 

The $L$-dependence of the single magnon energy for open chains has been 
found to behave as $1/L$ for medium chain length and as 
$\Delta + O(1/L^2)$ for long chain length\cite{qin0}.  In this paper we 
will show that those two behaviors for different range of lengths have 
a unified expression $\sqrt{\Delta^2+v^2\sin^2{\frac{\pi} {L+2}}}$. 

With periodic boundary conditions (PBC) finite size corrections to the 
groundstate and single magnon energies are exponentially small but, as 
we show, the lowest 2-magnon state (with S=2) has energy:
\begin{equation}
E_{2P}-E_{0P}=2\sqrt{\Delta^2+v^2\sin^2{\pi \over L-2a}}.
\label{EP}
\end{equation}
We will use the physical quantities obtained for S=1 spin chains in 
previous studies $\Delta=0.4105$ and $v=2.49$.   Note that the Lorentz
invariant dispersion relation $E=\sqrt{\Delta^2+(vk)^2}$ has been 
modified by the replacement:
\begin{equation}
k\to \sin k.
\end{equation}
(Throughout this paper we use wave-vectors shifted by $\pi$ so that the 
minimum energy single magnon excitation occurs at $k=0$.) This 
modification only affects the expansion in $1/L$ at $O(1/L^4)$.  To 
order $1/L^2$, this is just the energy of 2 free magnons with 
wave-vectors $\pm \pi /L$.  The fact that these wave-vectors are 
non-zero and different reflects the hard-core boson approximation in 
which the wave-function is approximated by a free fermion (Bloch) 
wave-function, multiplied by a sign function to correct the statistics. 
The antisymmetric nature of the Bloch wave-function requires the use of 
two different wave-vectors.  Periodic boundary conditions then requires 
them to be odd multiples of $\pm \pi /L$.  $a$ is a new parameter which 
we introduce, the magnon-magnon scattering length.  This  leads to an 
$O(1/L^3)$ correction:
\begin{equation}
E_{2P}-E_{0P} \approx 2\Delta 
	+ {1\over \Delta}\left({v\pi \over L}\right)^2
	+ \left({4a\over \Delta L}\right)
		\left( {v \pi\over  L}\right)^2 
	+ O(1/L^4).
\end{equation}

The replacement of $k$ by $\sin k$ is not derived here but is just an 
empirical improvement which reflects the presence of a lattice.  In 
particular, if we ignore $\Delta$ this gives the standard spin-wave 
theory formula. It gives a fairly good fit to the magnon dispersion 
relation over a large range of $k$\cite{tak} although it fails badly 
near $k=\pi$ (that is $k=0$ before the wave-vector shift).

For the case of open boundary conditions (OBC), we derive the following 
formulas for the L-dependence of the energies of the lowest energy 
states of spin S=0,1,2,3.
\begin{eqnarray}
\label{scal}
E_0&=&e_0(L-1)+\Delta_b-{{3J_{eff}}\over{4}}\exp(-{{L-1}\over{\xi}}),
	\nonumber \\
E_1&=&e_0(L-1)+\Delta_b+{{J_{eff}}\over{4}}\exp(-{{L-1}\over{\xi}}),
	\nonumber \\
E_2-E_1&=&\sqrt{\Delta^2+v^2\sin^2{\frac{\pi}{L-2a_b}}},\\
E_3-E_1&=&\sqrt{\Delta^2+v^2\sin^2{\frac{\pi}{L-2a_b-a}}} +
	\sqrt{\Delta^2+v^2\sin^2{\frac{2\pi}{L-2a_b-a}}}. \nonumber
\end{eqnarray}
We use the previously obtained parameters: $e_0=-1.401484$ and 
$\Delta_b=-0.193166$.  The exponentially small terms in $E_{0,1}$ arise 
from the interaction between the effective S=1/2 edge excitations.  
$E_2-E_1$ is the single magnon energy.  Note that it is approximately 
$\sqrt{\Delta^2+(vk)^2}$ with $k=\pi /(L-2a_b)$ up to corrections of 
$O(1/L^4)$.  This is simply the energy of a free massive particle in a 
box with an interaction at the boundaries producing a scattering length 
$a_b$.  The formula for $E_3-E_1$ gives the lowest energy two magnon 
state.  Note that the second wave-vector occurring here is twice as 
large.  This reflects the hard-core boson approximation.  The 
appropriate value of this second wave-vector is again determined from 
consideration of the boundary interactions and contains the same 
boundary scattering length $a_b$.  The same inter-magnon scattering 
length, $a$, appears as in the periodic case.  These formulas are only 
expected to be completely correct up to $O(1/L^3)$.  The same empirical 
replacement of $k$ by $\sin k$ has been made.  

The magnetization per unit length, at $T=0$, $L=\infty$, close to the 
critical field is given by:
\begin{equation}
M/L={1\over \pi v}\left[
	\sqrt{2\Delta (H-\Delta)}
	-{8\Delta a\over 3\pi v}(H-\Delta )
	\right].
\label{mag}
\end{equation}
(We adopt units where $g\mu_B=1$.) 
The square root singularity was first proposed by Tsvelik,\cite{tsv} 
using an approximate fermionic representation for the S=1 chain.  It 
was later argued\cite{aff2} to be the universal behavior of bosons 
with repulsive interactions, and therefore to be an exact result for 
integer spin chains.  The linear correction to this formula was derived 
by Okunishi et al.\cite{okun} recently by assuming a $\delta$-function 
interaction between magnons, $c\delta (x_i-x_j)$, with $a$ replaced by 
\begin{equation}
a\to -v^2/\Delta c .
\label{atoc}
\end{equation}
We argue here that this formula obtains for {\it any} short range 
interaction (which does not produce boundstates), the details of the 
interaction determining the scattering length.  

The scattering length 
is defined by the behavior of the (symmetric) scattering phase shift in
the limit of zero wave-vector.  For a symmetric wave-function 
describing the relative motion of 2 particles, the long distance 
behavior is written in terms of the phase shift, $\delta (k)$ as:
\begin{equation}
\psi (x)\to \sin [k|x|+\delta (k)].
\label{PSdef}
\end{equation}
As $k\to 0$, for general short range potentials with no boundstates in 
the limit $k\to 0$:
\begin{equation}
\delta (k)\to -ak.
\end{equation}
We note that the scattering length can be positive or negative.  An 
infinite hard core potential gives a positive scattering length equal 
to the core size.  On the other hand a repulsive $\delta$-function 
potential gives a negative scattering length given by Eq. (\ref{atoc}). 

We also derive the generalization of Eq. (\ref{mag}) at low temperature 
T and large 
L in order to fit recent Monte Carlo results.\cite{kash}  We obtain a 
consistent value of $a$ of about $-0.34\xi$ from all three fits, as 
mentioned above. However, using a product wave-function renormalization
group method to calculate the magnetization, Okunishi et al. obtained a 
considerably larger value, $a\approx -0.54\xi$, which is closer the 
prediction of the NL$\sigma$M.  From the viewpoint of the 
renormalization group treatment of the one dimensional Bose 
condensation transition,\cite{sachdev} we may regard $a$ as the leading irrelevant 
coupling constant.  Although its value is non-universal, many different 
quantities can be expressed in terms of it.  Thus it plays a similar
role to one over the Kondo temperature in the Kondo problem.

The boundary scattering length, $a_b$, resulting from the interaction 
of bulk magnons with the chain end and the effective S=1/2 degree of 
freedom residing there, is found to have a value of approximately -1 
lattice constants.  

In the next section we discuss the simpler case of periodic boundary 
conditions, deriving the wave-functions and energies for dilute bosons 
with short range interactions in terms of the scattering length and 
obtaining the scattering length for the NL$\sigma$M. We compare 
Eq. (\ref{EP}) to DMRG results to determine $a$.  In Sec. 3 we discuss 
the case of open boundary conditions, deriving Eqs. (\ref{scal}) and
comparing them to DMRG results, obtaining $a_b$ and a consistent value 
of $a$.  In Sec. 4 we discuss the magnetization, deriving 
Eq. (\ref{mag}) and its finite T and L generalizations, comparing to 
Monte Carlo results and again obtaining a consistent value of $a$.

\section{Dilute Bosons and the Non-linear $\sigma$ model}
Following the derivation of the NL$\sigma$M for large $S$ from the 
Heisenberg model\cite{hal1,aff0}, we first define 
\begin{equation}
\label{s2ph}
{\bf S}_{2i-1}=-S{\vec \phi}_i+{\bf l}_i,\;\;\;
{\bf S}_{2i}=S{\vec \phi}_i+{\bf l}_i,\;\;\;
i=1\;{\rm to}\; L/2,
\end{equation}
for a chain of $L$ sites. $\vec \phi$ and ${\bf l}$ represent the low 
energy Fourier modes of the spin operators with wave-vectors near 
$\pi$ and $0$ respectively. Starting from Eq.(\ref{sham}) we obtain a
continuum NL$\sigma$M Hamiltonian:
\begin{equation}
\label{nlsm}
{\cal H} = \int_0^{L} dx \left[ {1\over{S}}{\bf l}^2
+{S\over{4}}\left( { d{\vec \phi}\over{dx}}\right)^2 \right] +\ldots
\end{equation}
This gives a velocity of $2S$ and a coupling constant $g=2/S$.  
The non-linear constraints and commutation relations give the 
corresponding Lagrangian:
\begin{equation}
{\cal L}={1\over 2g}\partial_\mu \vec \phi \cdot 
			\partial^\mu \vec \phi ,
\end{equation}
with the constraint $\vec \phi^2=1$.  (We have set the effective 
``velocity of light'' $v$ to one.)  The spectrum of this field theory 
is known to consist of only a triplet of massive bosons, created by the 
fields, $\vec \phi$, and their multi-particle scattering states.  There 
are no boundstates.  The boson mass is exponentially small in the 
coupling constant, $g$; it results from non-perturbative effects.  A 
simpler Lagrangian which is expected to have qualitatively similar 
physics is the $\phi^4$ model in which the constraint on $\vec \phi$ 
is relaxed:
\begin{equation}
{\cal L}={1\over 2}\partial_\mu \vec \phi \cdot 
			\partial^\mu \vec \phi
	-{\Delta^2\over 2}\vec \phi ^2
	-{\lambda \over 4}\vec \phi^4.
\end{equation}
This model also has a triplet of massive bosons and presumably no 
boundstates for $\lambda >0$ (repuhlusive interactions).  Of course, the 
details of the boson interactions will be somewhat different in the two 
models.  However, the single boson energy, $\sqrt{\Delta^2+v^2k^2}$, 
for $\Delta$ the renormalized mass, is independent of the interactions 
and simply reflects Lorentz invariance and the stability of the single 
boson excitation. 

We now consider the low energy states of  a dilute gas of bosons, all 
with $S^z=+1$ such that the average spacing is much greater than the 
Compton wavelength of the boson, $\xi = v/\Delta$.   A great deal of 
universality occurs in this limit.  In particular since the bosons in 
these states have small wave-vectors, $k<<\xi^{-1}$, the usual 
non-relativistic approximation to the dispersion relation is adequate:
\begin{equation} 
\epsilon (k)\approx \Delta + {v^2k^2\over 2\Delta}.
\label{NLI} 
\end{equation} 
This implies that the behavior will not depend on the Lorentz 
invariance of the underlying field theory.  This is an important point 
because the S=1 chain is only approximately Lorentz invariant.  
Nevertheless, it contains stable bosonic magnons whose dispersion 
relation is given by Eq. (\ref{NLI}) for small $k$.  (Note that we have 
shifted the wave-vectors in the spin chain problem by $\pi$.)  This 
follows from the assumption that the dispersion relation is an even 
function of $k$ analytic near $k=0$ and defines the parameters $\Delta$ 
and $v$.  Due to spin conservation, the low energy  states under 
consideration are completely stable against decaying into states 
involving bosons of other spin polarizations, so they may be integrated 
out.  At low energies, the effect of all processes involving creation 
and annihilation of virtual particles, as given, for example by the 
Feynman diagram expansion of the $\phi^4$ theory, may be boiled down to 
a simple non-relativistic effective Hamiltonian, written in first 
quantized notation as: 
\begin{equation} 
H_N=-{1\over 2m}\sum_{i=1}^N{d^2\over dx_i^2} 
	+\sum_{i,j}V(x_i-x_j). 
\label{HNR}
\end{equation} 
Here $m\equiv \Delta /v^2$ (the Einstein relation) and $V$ is some effective short
range interaction.  We emphasize that the exact form of $V$ is unknown in all cases
except for the weak coupling limit of the $\phi^4$ model where it is simply a
repulsive $\delta$ function.  We {\it do} know that it is such as not to produce any
boundstates and we expect its range to be roughly $\xi$.  We may indirectly infer its
properties, in the case of the NL$\sigma$M, from the  low energy S-matrix which is
known exactly.  Nothing is known, a priori about the effective potential for the S=1
chain, but we may indirectly deduce some of its properties from numerical
simulations.  Importantly,  in the dilute limit the spectrum only depends very weakly
on the detailed form of $V$.  

In this dilute limit we may construct the many body wave-functions following
the construction of Lieb and Liniger for the $\delta$-function
Bose gas.\cite{lieb}  This follows because 
we may ignore the possibility of more than 2 particles being within a distance
$\xi$ of each other simultaneously so we only have to consider 2-particle
scattering processes.  In the most likely case where all particles are
far from each other compared to $\xi$, we may write the N-body wave-function
as a sum of plane waves.  Considering the case $x_1<x_2\ldots <x_N$, we write:
\begin{equation}
\psi (x_1,\ldots x_N)\approx \sum_PA(P)P\exp (i\sum_{j=1}^Nk_jx_j),
\label{MBWF}\end{equation}
for some set of wave-vectors, $k_j$.  Here $P$ permutes the $k_j$'s and
the sum is over all permutations.  $\psi$ is determined for other 
orderings of the $x_i$ from the symmetry of the wave-function following 
from Bose statistics.  By considering what happens when 2 particles 
approach each other, we can see that
\begin{equation}
A(Q)=-A(P)e^{i2\delta [(k_i-k_j)/2]}.
\label{a(P)}
\end{equation}
Here the two permutations $P$ and $Q$ differ by the interchange of only 
1 pair, $i$ and $j$ and $\delta (k)$ is the even channel phase shift 
defined by the 2-body problem with potential $V(x)$.  The Hamiltonian 
for this problem, where we consider only the relative motion of a pair 
of particles, defining $x=x_1-x_2$, is:
\begin{equation}
H = -{1\over m}{d^2\over dx^2}+V(x).
\end{equation}
(Note that the reduced mass, $m/2$ occurs here.)  The even 
wave-functions, at distances $|x|>>\xi$ take the form of 
Eq. (\ref{PSdef}), defining $\delta (k)$.  In general, the coefficients 
$A(P)$ in Eq. (\ref{MBWF}) are just products of the factor in 
Eq. (\ref{a(P)}) over all all elementary permutations corresponding to 
$P$.  By considering again, a region where all the particles are far 
apart, so $V\approx 0$, we see that:
\begin{equation}
E\approx {1\over 2m}\sum_jk_j^2.
\end{equation}
Periodic boundary conditions then determine the $k_j$'s, which must all
be different from each other, from the N equations:
\begin{equation}
(-1)^{N-1}e^{-ik_jL}=\exp\left(
		i\sum_{s=1}^N2\delta [(k_s-k_j)/2]
			\right),\ \ \hbox{all j}.
\label{detk}
\end{equation}
We emphasize that this result only holds in the dilute limit when all 
the $k_j$'s are small.  Thus we are only interested in the phase shift 
for small $k$.  It is easy to see that 
\begin{equation}
\delta (k)\to -ak,
\label{delapp}
\end{equation}
for some constant $a$ at small $k$, $k<<1/\xi$.  This follows from 
observing that in this limit, in the region $\xi <<|x|<<1/k$, the 
wave-function obeys approximately the zero energy, zero potential 
Schroedinger equation:
\begin{equation}
-{1\over m}{d^2\over dx^2}\psi (x)=0.
\end{equation}
The even solutions have the form:
\begin{equation}
\psi \propto |x|-a.\label{psias}
\end{equation}
Matching this with $\sin [k|x|+\delta (k)]$ at large $|x|$ then
determines $\delta \approx -ak$.  By analogy with the standard 
definition in three dimensions, we refer to $a$ as the scattering 
length.  We discussed it briefly in Sec. 1.  We generally expect it to 
be of $O(\xi)$.  

It is perhaps worth remarking here that, with our definition of phase
shift, it takes the value $\pi /2$ for all $k$ for $V=0$ since the
symmetric wave-functions are then $\cos kx$.  In this non-interacting 
case Eq. (\ref{psias}) is not valid. The fact that $\delta (k)\to 0$ at 
$k\to 0$ for {\it any} short range repulsive interaction, no matter how 
weak, represents a sort of infrared singularity of one dimensional 
single particle quantum mechanics.  It is basically this singularity 
which is responsible for the fact that, in the dilute limit, the many 
body wave-function for interacting bosons reduces to that of free 
fermions times an antisymmetric sign function.\cite{aff2}  A related 
observation is that $\delta (k)=0$ implies the $k_j$ must all be 
different since otherwise the wave-function of Eqs. (\ref{MBWF}), 
(\ref{a(P)}) would vanish.

\begin{figure}[ht]
 \epsfxsize=3.3 in\centerline{\epsffile{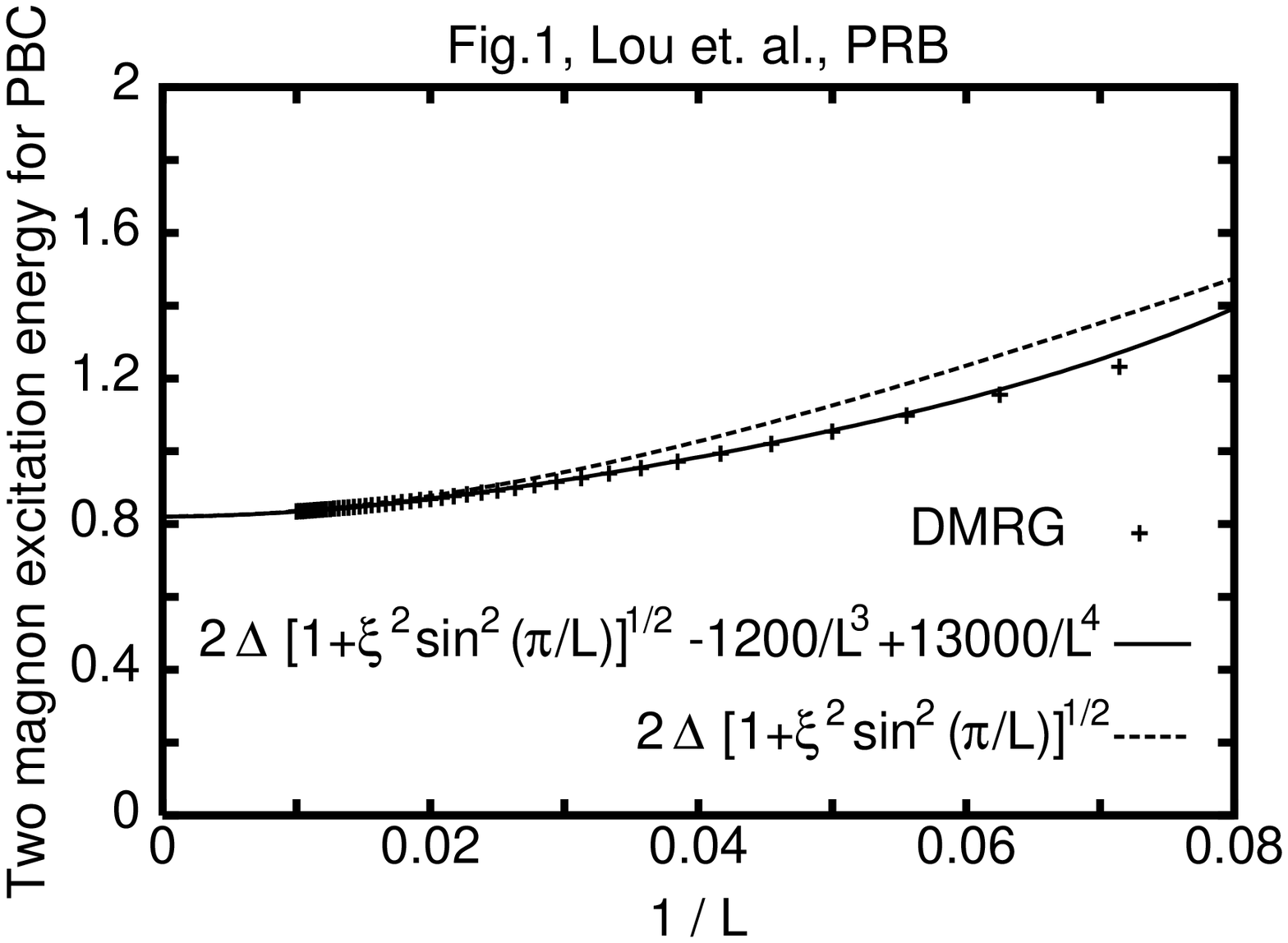}}
\vspace{0.5cm}
\caption[]{
Excitation energy of two magnons state with PBC, 
$E_{2P}-E_{0P}$. }
\label{E2-E0P}
\end{figure}

In general, it is non-trivial to solve Eq. (\ref{detk}), requiring 
numerical methods.  However, in the small $k$ limit, where 
Eq. (\ref{delapp}) applies, it is easy.  Since the phase shift for the 
$\delta$-function potential and the hard core potential have this form 
at low $k$, we simply recover the low density, low $k$ limit of those 
problems.\cite{lieb}  In that limit,  Eq. (\ref{detk})  can be written: 
\begin{equation} 
k_j(L-Na)+a\sum_sk_s=\pi n_j,\ \  \hbox{all j},
\label{kj}
\end{equation} 
where the integers, $n_j$ are all even, for N odd and all odd for N 
even. In the case $N=2$ the lowest energy solution of 
Eq. (\ref{kj}) is: 
\begin{equation} 
k_1=-k_2={\pi \over L-2a}
\label{kPBC}
\end{equation} 
and the energy is: 
\begin{equation} 
E\approx {1\over m}\left( {\pi \over L-2a}\right)^2
 \approx {1\over m} \left({\pi \over L}\right)^2
	+{4a\over mL}\left({\pi \over L}\right)^2.
\label{EPBC} 
\end{equation}
This is fitted to our DRMG results for the excitation energy of the 
lowest state with $S^z=2$ with PBC in Fig. \ref{E2-E0P}.  A $1/L^3$ is 
clearly discernible and gives an estimate of the inter-magnon 
scattering length: 
\begin{equation}
a\approx -0.34 \xi 
\end{equation}  
We emphasize that a negative scattering length does not imply an 
attractive interaction.  In particular, as mentioned in Sec. 1, a 
repulsive $\delta$-function interaction gives $a<0$.  

The exact S-matrix has been conjectured for the O(n) NL$\sigma$M based 
on factorization of m-particle scattering and various consistency
conditions and checked against the $1/n$ expansion.\cite{zam}  The three
possible states of a single magnon are labeled by a vector index
$i=1$, $2$, $3$ so that $|A_i>\propto \phi^i|0>$.  O(3) symmetry
then implies that the S-matrix takes the form:
\begin{eqnarray}
_{ik}S_{jl}&\equiv& <A_j(p_1')A_l(p_2'),\hbox{out}
		|A_i(p_1)A_k(p_2),\hbox{in}> \label{exS}\nonumber \\
&=&\delta (p_1-p_1')\delta (p_2-p_2')
		[\delta_{ik}\delta_{jl}\sigma_1(\theta ) 
			+\delta_{ij}\delta_{kl}\sigma_2(\theta )
			+\delta_{il}\delta_{jk}\sigma_3(\theta )]
	+(i\leftrightarrow k,p_1\leftrightarrow p_2).
\end{eqnarray}
Here the momenta of the particles are labeled by the rapidities 
$\theta_a$:
\begin{equation} 
vp_a=\Delta \sinh \theta_a, 
\end{equation}
and in Eq. (\ref{exS}) the $\sigma_i$'s are expressed as functions of
\begin{equation} 
\theta \equiv \theta_1-\theta_2.
\end{equation}
These functions are given by:
\begin{eqnarray}
\sigma_1(\theta )&=&
	{2\pi i\theta \over (\theta+\pi i)(\theta -2\pi i)}, \nonumber\\
\sigma_2(\theta )&=&
	{\theta (\theta -\pi i)\over (\theta+\pi i)(\theta -2\pi i)},
				\nonumber \\
\sigma_3(\theta )&=&
	{-2\pi i(\theta -\pi i)\over (\theta+\pi i)(\theta -2\pi i)}.
\end{eqnarray}
The S-matrix for scattering of 2 spin-up particles is obtained by using:
\begin{equation}
|A_{+}(p)>=(|A_{1}(p)>+i|A_{2}(p)>)/\sqrt{2},
\end{equation}
giving:
\begin{eqnarray}
&&<A_+(p_1')A_+(p_2'),\hbox{out}|A_+(p_1)A_+(p_2),\hbox{in}>
			\nonumber \\
&&=-\delta (p_1-p_1')\delta (p_2-p_2')
	{1+i\theta /\pi \over 1-i\theta /\pi }
   -\delta (p_2-p_1')\delta (p_1-p_2')
	{1-i\theta /\pi \over 1+i\theta /\pi }.
\end{eqnarray}
This determines the phase shift:
\begin{equation}
e^{2i\delta [(p_1-p_2)/2]}={1+i\theta /\pi \over 1-i\theta /\pi}.
\end{equation}
Now letting $p_1-p_2\to 0$, we obtain the exact scattering length of
the NL$\sigma$M, for 2 spin up magnons:
\begin{equation}
a_{NL\sigma M}=-2v/\Delta \pi =-2\xi /\pi,
\end{equation}
roughly twice as large as the value determined numerically for the S=1 
Heisenberg model.   

This calculation can also be repeated for 2 magnons in a state of total
spin S=1 or S=0.  The S=1 states are produced from the groundstate by
the low wave-vector components of the lattice spin operators.  We obtain
the scattering lengths for spin-S from the NL$\sigma$M:
\begin{eqnarray} 
a^2&=&-2\xi /\pi \nonumber \\
a^1&=&-\xi /\pi \nonumber \\
a^0&=&\xi /\pi .
\end{eqnarray}
For the S=0 case, where the wave-function is again symmetric with 
respect to the spatial co-ordinates, Eqs. (\ref{kPBC}) and (\ref{EPBC}) 
still apply with $a$ replaced by $a^0$.  The S=1 case is a bit 
different because now the wave-function must be anti-symmetric with 
respect to the spatial co-ordinates since it is anti-symmetric with 
respect to the spin indices.  In this case we obtain the wave-vectors 
for the lowest energy states:
\begin{equation}
k_1\approx 0,\ \  k_2\approx \pm 2\pi /(L-a^1),
\end{equation}
and the energy:
\begin{equation}
E_{1P}'-E_{0P}\approx 2\Delta 
	+{1\over 2m}\left({2\pi \over L}\right)^2
	+{1\over m}\left({2\pi \over L}\right)^2{a^1\over L}.
\label{EPS1}
\end{equation}

\section{Open boundary conditions}
\begin{figure}[ht]
 \epsfxsize=3.3 in\centerline{\epsffile{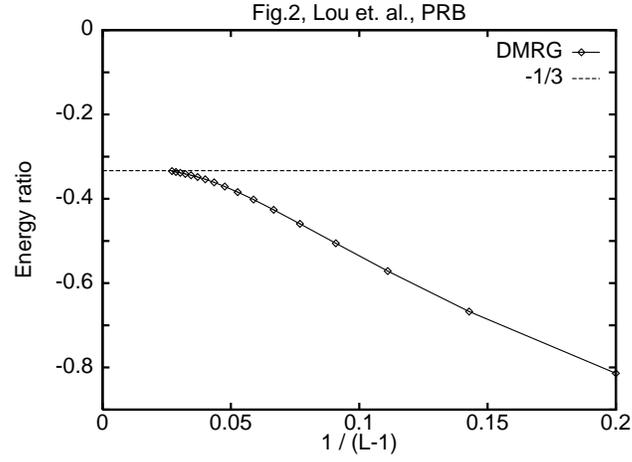}}
\vspace{0.5cm}
\caption[]{
Ratio of coupling energy of two edge $1/2$ spins in ground state 
and edge excited state. Here energy ratio 
$\frac{E_1-e_0(L-1)-\Delta_b}{E_0-e_0(L-1)-\Delta_b}$ vs. $1/(L-1)$ 
is plotted. Dashed line is $-\frac{1}{3}$.
}
\label{E1/E0}
\end{figure}

\begin{figure}[ht]
 \epsfxsize=3.3 in\centerline{\epsffile{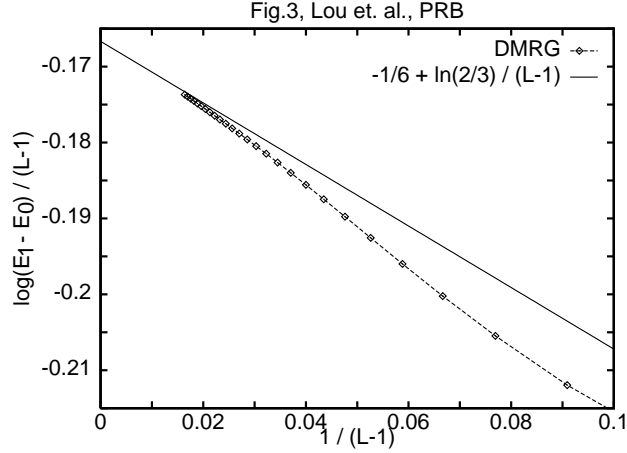}}
\vspace{0.5cm}
\caption[]{
Coupling constant for edge spins.  Here $\ln (E_1-E_0)/(L-1)$ vs.
$1/(L-1)$ is plotted.  The straight fitting line is 
$\ln (E_1-E_0)/(L-1)=-1/\xi+c^{\prime}/(L-1)$, where
$\xi=6$ and $c^{\prime}$=2/3 is taken from Ref.[\onlinecite{qin0}].
}
\label{E1-E0}
\end{figure}

While a periodic integer spin chain has an excitation gap which depends 
only weakly on L, in the case of OBC there are low lying edge 
excitations. These correspond to effective S=1/2 spins localized at the 
two ends of the chain which are coupled to the bulk degrees of freedom. 
An effective field theory is given by the NL$\sigma$M with Neumann 
boundary conditions,
\begin{equation}
d\vec \phi /dx=0, \ \  (x=0,L),
\end{equation}
coupled to the S=1/2 spins, ${\bf S}_1$ and ${\bf S}_2$.  In general, 
one expects a coupling to both the staggered and uniform magnetization 
density at the chain ends:\cite{sor3}
\begin{equation}
H=H_{\hbox{bulk}}+\lambda_s[\vec \phi (0)\cdot {\bf S}_1 
				+(-1)^L\vec \phi (L)\cdot {\bf S}_2]
		+\lambda_u[{\bf l}(0)\cdot {\bf S}_1 
				+{\bf l}(L)\cdot {\bf S}_2].
\end{equation}
Integrating out the bulk field, $\vec \phi$ gives an effective 
exponentially small interaction, between the end spins proportional to
$\lambda_s^2$. 
\begin{equation}
J_{eff}\propto \lambda_s^2e^{-L/\xi}.
\end{equation}
This leads to the splitting between the lowest singlet and triplet 
states in Eq. (\ref{scal}) which can simply be thought of as formed 
from the two S=1/2 end excitations. The singlet (triplet) state of the 
two S=1/2's has lowest energy for L even (odd). (We only consider L 
even here.) We check the formulas in Eq. (\ref{scal}) for $E_0$ and 
$E_1$ in Figs. \ref{E1/E0} and \ref{E1-E0}.  Because $J_{eff}$
 is exponentially small, it becomes smaller than
our numerical error for the longest chain lengths that we have studied.
Therefore, we do no show data for these chain lengths in 
Figs. \ref{E1/E0} and \ref{E1-E0} since it would be meaningless.  This
problem does not occur for the other energy differences that we
consider which only
involve finite size effects scaling as powers of $1/L$. 

The lowest $S=2$ state is obtained by polarizing the two end spins and 
then adding one $S^z=1$ magnon.  When the end spins are polarized, the 
$\lambda_u$ term produces an effective boundary potential energy acting 
on the magnons.  We may describe this by adding an effective boundary 
potential for the magnons.  Thus the non-relativistic N-body 
Hamiltonian of Eq. (\ref{HNR}) gets modified to:
\begin{equation}
H_{N}=-{1\over 2m}\sum_{i=1}^N{d^2\over dx_i^2} 
	+\sum_{<i,j>}V(x_i-x_j) 
	+\sum_{i=1}^N[V_b(x_i)+V_b(L-x_i)],
\end{equation}
where $0\leq x_j\leq L$ with the Neumann boundary condition:
\begin{equation}
\partial_j\psi (x_1,x_2,\ldots x_N)=0, \ \  (x_j=0,L).
\end{equation}
We expect $V_b(x)$ to be some short range function (with range of 
$O(\xi )$).  It is not a priori clear whether $V_b$ is attractive or 
repulsive.  However, the absence of boundary boundstates that is 
evident from earlier numerical work\cite{sor4} indicates that it is 
repulsive.  

\begin{figure}[ht]
 \epsfxsize=3.3 in\centerline{\epsffile{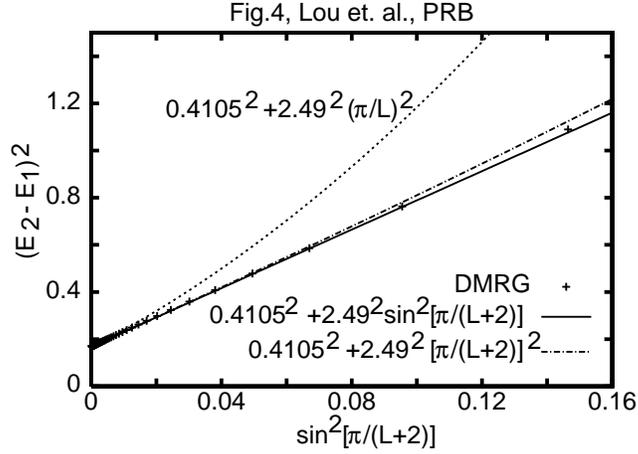}}
\vspace{0.5cm}
\caption[]{
Single magnon energy with OBC.  Here $(E_2-E_1)^2$ vs.
$\sin^2{\frac{\pi}{L+2}}$ is plotted.  The boundary scattering length, 
$a_b\approx -1$ is obtained.  The straight fitting line shows a better 
fit at short chain lengths results from the replacement of $k$ by 
$\sin k$.  
}
\label{E2-E1O}
\end{figure}

For low energy multi-magnon states only the scattering length $a_b$,
produced by $V_b$ is relevant to the spectrum.  This has the effect of 
modifying the low energy single magnon wave-functions, at distances 
large from the boundaries compared to $\xi$, to the form:
\begin{equation}
\psi (x)\approx \sin [k(x-a_b)] \propto \sin [k(L-a_b-x)].
\end{equation}
Consistency of these expressions quantizes the wave-vectors:
\begin{equation} 
k_j=\pi j/(L-2a_b),\ \  j=1,2,3\ldots .
\end{equation}
The effective chain length becomes $L-2a_b$.  It is now clear that our 
results are quite robust against variations in how the effective 
Hamiltonian is written.  For instance, it doesn't matter whether we 
impose the Neumann b.c. at $x=0$ or $x=1$.  Such differences can be 
absorbed into the scattering length.  We also note that $<S^z_j>$ 
exhibits period $\pi$ oscillations upon which are superimposed long 
wavelength variations which can be interpreted in terms of the magnon 
wave function.\cite{sor4}  The boundary scattering length can be 
determined from the additional energy to add one $S^z=1$ magnon to the 
polarized end spins, giving the third of Eqs. (\ref{scal}).  This gives 
a $1/L^3$ term:
\begin{equation}
E_2-E_1\approx \Delta + {1\over 2m}\left( {\pi\over L}\right)^2 
		+{2\over m}{a_b\over L} \left( {\pi\over L}\right)^2.
\end{equation}
Fitting to this expression, as shown in Fig. \ref{E2-E1O}, we obtain:
\begin{equation} 
a_b\approx -1.
\end{equation}
This is equivalent to imposing vanishing boundary conditions on a chain 
of length $L+2$.  We remark that there is no simple derivation that we 
know of for this result.  Presumably the value of $a_b$ depends on the 
(integer) magnitude of the spin and other details of the microscopic 
Hamiltonian.  We note that a somewhat better fit is obtained in 
Fig. \ref{E2-E1O} by the replacement $k\to \sin k$.
  
\begin{figure}[ht]
 \epsfxsize=3.3 in\centerline{\epsffile{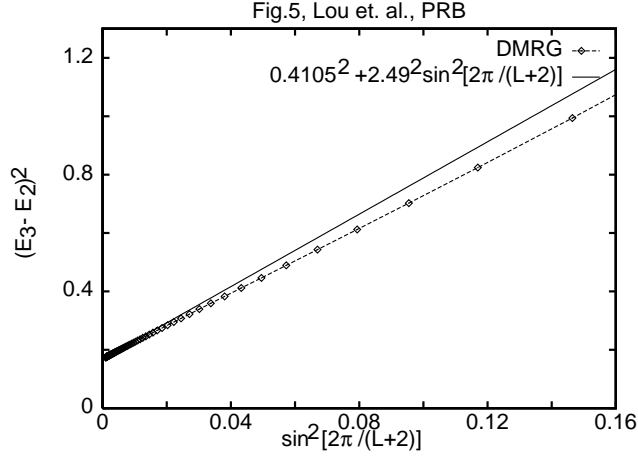}}
\vspace{0.5cm}
\caption[]{
Two magnon excitation behavior for $S=1$ open chain.  Here 
$(E_3-E_2)^2$ vs.  $\sin^2{\frac{2\pi}{L+2}}$ is plotted.
The straight line is $(E_3-E_2)^2=\Delta^2+v^2\sin^2{\frac{2\pi}{L+2}}$
without any behavior of magnon-magnon interaction included.
}
\label{E3-E2O}
\end{figure}

\begin{figure}[ht]
 \epsfxsize=3.3 in\centerline{\epsffile{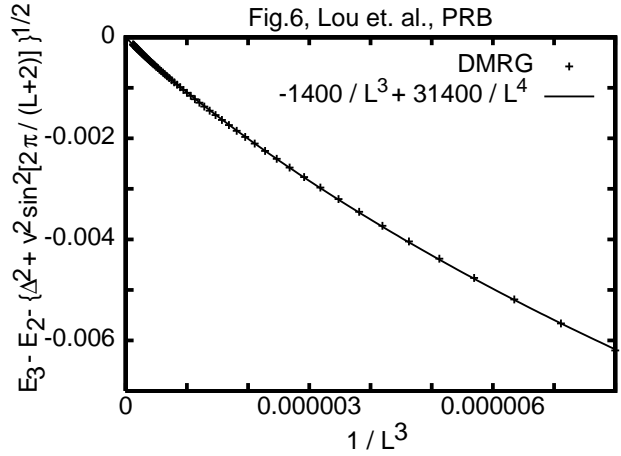}}
\vspace{0.5cm}
\caption[]{
Two magnon excitation energy for OBC.  Here 
$E_3-E_2-\sqrt{\Delta^2+v^2\sin^2{\frac{2\pi}{L+2}}}$ vs. $1/L^3$ 
is plotted.  The fitting curve goes to zero linearly as $1/L^3\to 0$.  
The coefficient of the cubic term determines the scattering length, 
$a\approx -0.32 \xi$.
}
\label{E3-E1O}
\end{figure}

We now consider the spin-polarized two-magnon wave-function in the 
presence of the polarized end spins, with $S^z=3$.  The effect of the 
boundary potential is simply to fix the effective size of the chain at 
$L-2a_b\equiv L'$ with an effective vanishing boundary condition.  The 
2-magnon wave-function, with wave vectors $k_1$ and $k_2$ can then be 
written in terms of the magnon-magnon phase shift, $\delta (k)$.  This 
wave-function is made from linear combinations of the periodic 2-magnon 
wave functions discussed in Sec. 2 with wave-vectors $(k_1,k_2)$
$(k_1,-k_2)$, $(-k_1,k_2)$ and $(-k_1,-k_2)$.  Two different
magnon-magnon phase shifts occur:
\begin{equation} 
\delta_\pm \equiv \delta [(k_1\pm k_2)/2].
\end{equation}
for $x_1<x_2$, this wave-function, constructed to obey
$\psi (0,x_2)=0$, is given by:
\begin{eqnarray}
&&\psi(x_1,x_2)=e^{i(k_1x_1+k_2x_2)}-e^{i[(k_2x_1+k_1x_2)+2\delta_-]}
   -e^{i(-k_1x_1+k_2x_2)} +e^{i[(k_2x_1-k_1x_2)-2\delta_+]}\nonumber \\
&&+[-e^{i(k_1x_1-k_2x_2)}+e^{i[(-k_2x_1+k_1x_2)+2\delta_+]}
   +e^{-i(k_1x_1+k_2x_2)}-e^{-i[(k_2x_1+k_1x_2)-2\delta_-]}]
e^{2i(\delta_--\delta_+)}.
\label{OBCWF}
\end{eqnarray}
The wave-function for $x_2<x_1$ is obtained by interchanging $x_1$ and 
$x_2$ in order to enforce the symmetry required by Bose statistics.  
Imposing $\psi (x_1,L')=0$ then requires:
\begin{eqnarray}
e^{ik_2L'}&=&e^{-ik_2L'+2i(\delta_--\delta_+)}\nonumber \\
e^{ik_1L'+2i\delta_-}&=&e^{-ik_1L'-2i\delta_+}.
\end{eqnarray}
Now using the small $k$ approximation to $\delta (k)$ gives the 
conditions:
\begin{eqnarray}
2k_2(L'-a)k_2&=&2\pi n_2\nonumber \\
2k_1(L'-a)k_1&=&2\pi n_1.
\end{eqnarray}
Thus the allowed wave-vectors are:
\begin{equation}
k_i={\pi n_i\over L-2a_b-a},
\label{kOBC}
\end{equation}
The lowest energy state has $n_1=1$, $n_2=2$.  Fig. \ref{E3-E2O}
shows that the second magnon has momentum $k_2=\frac{2\pi}{L}$
with $n_2=2$. [The $n_i$ must be 
different in order for the wave-function of Eq. (\ref{OBCWF}) not to 
vanish.] Setting $a_b=-1$, its energy is given by the last of 
Eq. (\ref{scal}).  As can be seen from Fig. \ref{E3-E1O}, good 
agreement with this formula is obtained with a value of the 
magnon-magnon scattering 
length:
\begin{equation}
a\approx -0.32 \xi.
\end{equation}
This appears consistent, within the numerical error, with the result
$a\approx -0.34\xi$ obtained with PBC.  

We also comment briefly on finite size energies of two magnon states 
with other spin quantum numbers.  This would certainly be simplest to 
study using PBC so that there are no end spins to worry about.  An 
alternative approach, following White and Huse,\cite{hus} would be to consider 
OBC but add extra S=1/2 variables at the edges of the system so as to 
cancel the effect of the edge excitations.  The boundary scattering 
length will depend on the details of the boundary couplings.  The 
contribution of the magnon-magnon interaction energy to the 2 magnon 
states can again be expressed in terms of the scattering length for 
the appropriate spin channel, $a^S$.  It can be seen that the 
wave-vectors are still given by Eq. (\ref{kOBC}) with the appropriate 
value of $a$ for both symmetric (S=0,2) and anti-symmetric (S=1) 
wave-functions.  Thus the energy, to $O(1/L^3)$ in all cases is given 
by:
\begin{equation}
E_{SO}-E_{0O}\approx 2\Delta +{5\over  2m}\left({\pi \over L'}\right)^2
	+{5\over  m}\left({\pi \over L'}\right)^2{a^S\over L}
\end{equation}
where $L'=L-2a_b$ and $a_b$ is the appropriate boundary scattering
length.

DMRG results on 2 magnon energies of various S were reported by White 
and Huse.  They found repulsive interaction energy for S=0,2 but 
attractive for S=1.  This is completely consistent with our approach.  
They did not use unequal wave-vectors for the 2 magnons in their 
definition of the interaction energy, in the even S case.  Thus their 
positive interaction energy is presumably of $O(1/L^2)$ and just 
reflects the effective Fermi statistics of the dilute Bose system, i.e. 
the requirement of unequal wave-vectors.  The attractive interaction 
energy in the S=1 channel corresponds to a negative $a^{1}$.  They only 
present data for L=60, but if we assume that the $1/L^3$ term is 
dominating at that value of L we extract a value, 
$a^1\approx -0.43\xi$, about 30\% larger in magnitude than the 
NL$\sigma$M prediction.  

\section{Magnetization}
We now consider wave-functions, with periodic boundary conditions, 
containing an arbitrary number, $N$, of magnons, assuming that the 
density and all wave-vectors are small compared to $1/\xi$.  To solve 
Eq. (\ref{kj}) for the $k_j$'s for a general low energy $N$-particle 
state we now use the fact that the density is low, $Na/L<<1$ and 
$a\sum_sk_s<<k_jL$.  Hence, in lowest order approximation:
\begin{equation}
k_j\approx k_{j0}\equiv \pi n_j/L.
\ \  (n_j\ \hbox{even for N odd}, n_j\ \hbox{odd for N even})
\end{equation}  
The dimensionless expansion parameter is $n a$ where $n$ is the
density, $N/L$.  Hence we may expand the $k_j$ in powers of $a$.  The
leading correction is $k_j=k_{j0}+\delta k_j$, where
\begin{equation}
\delta k_j={Na\over L}k_{j0}-{a\over L}\sum_sk_{s0}.
\end{equation}
The energy of this state is approximately:
\begin{equation}
E\approx {1\over 2m}\sum_j[k_{j0}^2+2k_{j0}\delta k_j]
   ={1\over 2m}\sum_jk_{j0}^2 
	+ {a\over 2mL}\sum_{<i,j>}(k_{i0}-k_{j0})^2.
\label{energy}
\end{equation}

\begin{figure}[ht]
 \epsfxsize=3.3 in\centerline{\epsffile{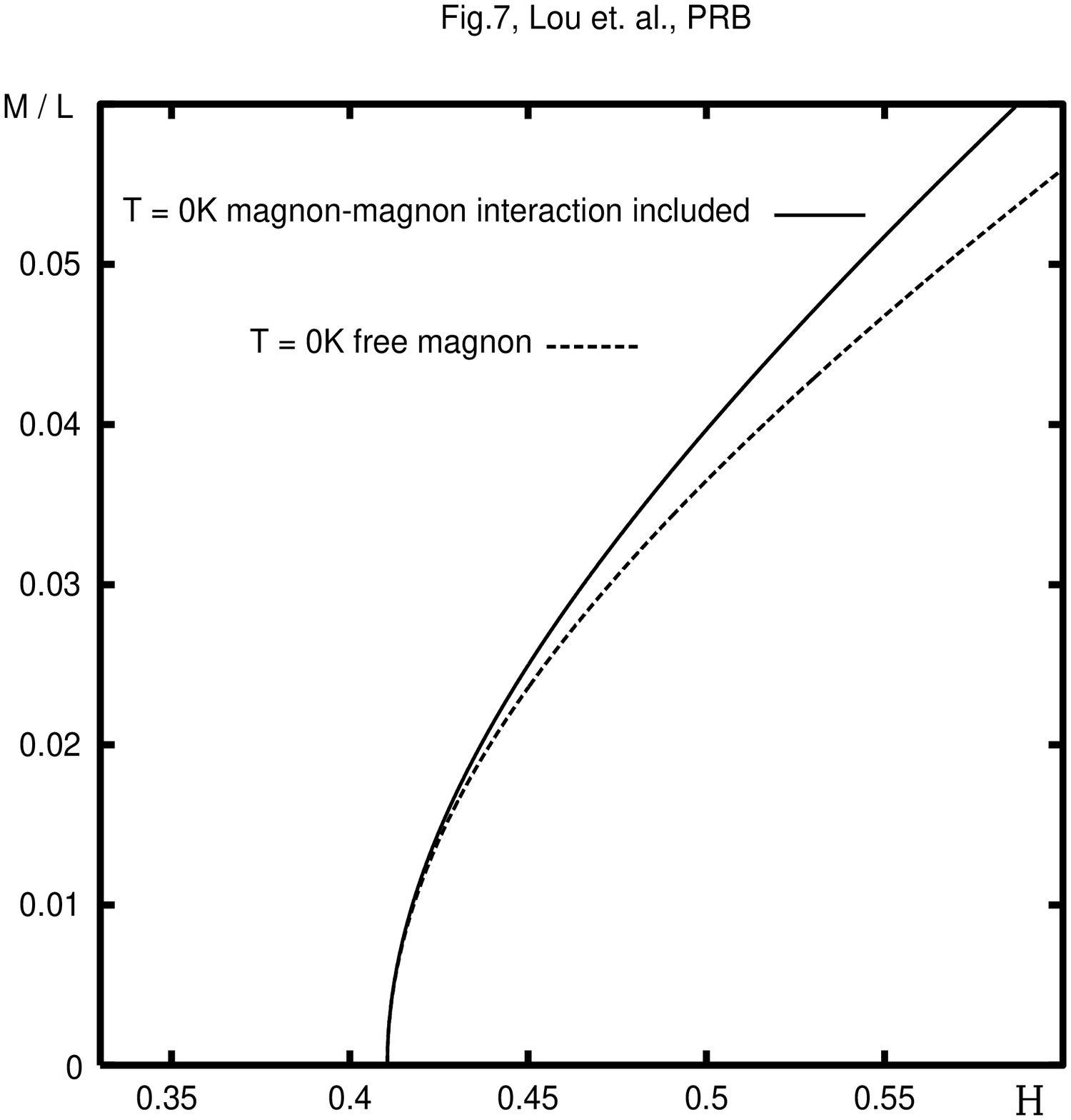}}
\vspace{0.5cm}
\caption[]{
Magnetization curve for $S=1$ chain near critical field $H_c=\Delta$.
(The exchange constant and $g\mu_B$ are set equal to 1.)
  For infinite length and at $T=0$,  we 
plot it as given in Eq.(\ref{magt}). Full line has included the leading 
order contribution of magnon-magnon interactions.  Dashed line is a 
reference line for comparison and it is for free hard core boson 
approximation.
}
\label{MT=0}
\end{figure}

We now consider the groundstate energy in the limit $L\to \infty$ with
a non-zero but small density, $n$.  The values of $k_{j0}$ occupy the
``Fermi sea'', $|k|<k_F$ where $k_F$ is determined from the density:
\begin{equation}
n=\int_{-k_F}^{k_F} {dk\over 2\pi}={k_F\over \pi}.
\end{equation}
The groundstate energy, to $O(a)$ is then:
\begin{equation}
E_0(n)/L=(\Delta -H)n
	+\int_{-k_F}^{k_F}{dk\over 2\pi}{k^2\over 2m}
	+{a\over 2m}\int_{-k_F}^{k_F}{dk\over 2\pi}
			\int_{-k_F}^{k_F}{dk'\over 2\pi}(k-k')^2.
\label{Eint}\end{equation}
Here we have included the Zeeman term, $-Hn$ in the energy; $H$ is the
applied magnetic field.  Performing the integrals gives:
\begin{equation}
E_0(n)/L=(\Delta -H)n +{\pi^2n^3\over 6m}
		+{a\pi^2n^4\over 3m}.
\end{equation}
This represents an expansion of the energy in powers of the density.  
The first term which depends on the interactions is the $O(n^4)$ term.  
Minimizing $E_0$ with respect to $n$ gives the density or magnetization 
per unit length:
\begin{equation}
M(H)/L=n\approx {\frac {1} {\pi v} }
	\left[\sqrt{2\Delta(H-\Delta)}-
	{\frac {8a\Delta} {3\pi v}}(H-\Delta)\right].
\label{magt}
\end{equation}
In Fig.\ref{MT=0} we plot $M/L$ vs. $H$ from Eq.(\ref{magt}) with and 
without the interaction ($a=-0.34\xi$ or $a=0$). The leading order 
correction due to the magnon-magnon interaction is obvious around 
magnetization $M/L=0.02$.

\begin{figure}[ht]
 \epsfxsize=3.3 in\centerline{\epsffile{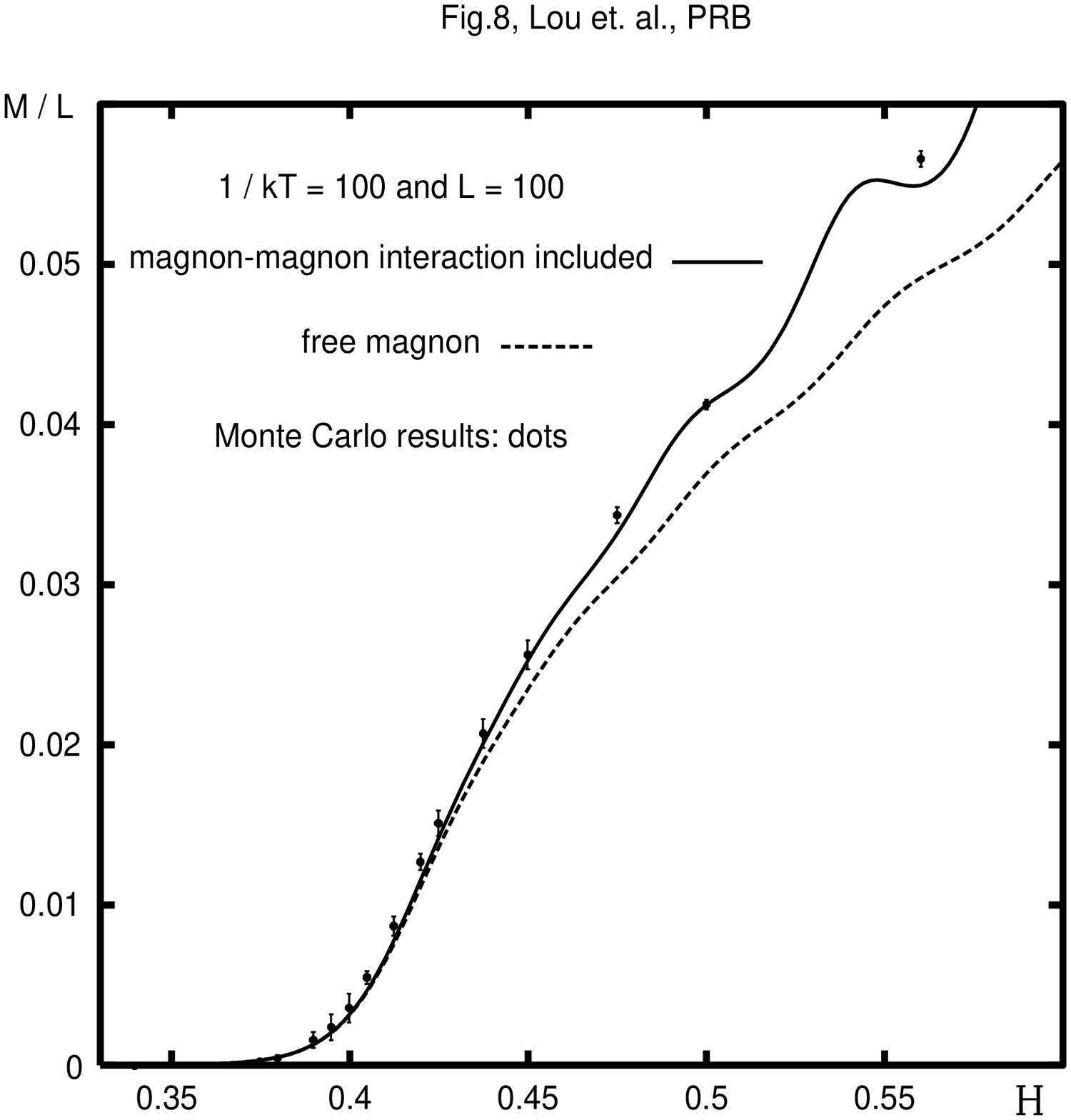}}
\vspace{0.5cm}
\caption[]{
Magnetization curve for $S=1$ chain near critical field $H_c=\Delta$,
with length $L=100$ and at temperature $kT = 1/100$. 
Full line has included the leading order contribution of magnon-magnon 
interaction.  Dashed line is a reference line for comparison and it is 
for free hard core boson approximation.  The dots are the Monte Carlo 
results\cite{kash}. 
}
\label{MC}
\end{figure}

This finite density correction can be generalized to finite 
temperature, $T$, as was observed by Okunishi\cite{okun2} in the 
special case of the $\delta$-function interaction.  The correction to 
the free energy of lowest order in density is given by including
thermal occupation numbers in the second term of Eq. (\ref{Eint}).
\begin{equation}
\Delta F =\frac {La }{ 2m}
\int \frac {dk_1dk_2}{4\pi^2}n_F(k_1)n_F(k_2)(k_1-k_2)^2.
\end{equation}
Note that it is the {\it Fermi} distribution function which appears, 
rather than the Bose function.  This simply follows from the condition 
that the $k_j$ should all be distinct so that there is an effective 
occupation number for each momentum which can be 0 or 1 only.  $n_F(k)$ 
is evaluated at finite $T$ and $H$, then the magnetization is obtained 
by the usual thermodynamic formula, $\partial F/\partial H=-M$.  In 
order to compare with recent Monte Carlo data on the magnetization for 
the S=1 chain, it is useful to also generalize our formulae to finite 
length, $L$ with periodic boundary conditions.  There is a slight 
subtlety in doing so because the allowed wave-vectors of the magnons 
alternate between $k=2\pi n/L$ (``even wave-vectors'') for an odd 
number of magnons and $k=2\pi (n+1/2)/L$ (``odd wave-vectors'') for an 
even number.  This follows from the sign change of the wave-function 
each time one magnon passes another one.  However, this is easily dealt 
with exactly by inserting appropriate factors of:
\begin{equation}
(-1)^N=e^{i\pi \sum_kn_k},
\end{equation}
into the partition function trace.  This effectively gives the chemical 
potential an imaginary part, essentially converting fermion occupation 
numbers into boson ones.  For $a=0$, the 
partition function is given by:
\begin{equation}
Z^0=(1/2)[Z^0_{Fe}-1/Z^0_{Be}+Z^0_{Fo}+1/Z_{Bo}^0 ].
\end{equation}
Here $Z^0_{Fe}$ denotes the partition function for free fermions with 
even wave-vectors; $Z^0_{Bo}$ denotes the partition function for bosons 
with odd wave-vectors, etc.  Note that {\it inverse} boson partition 
functions occur.  The thermal average of the second  term in 
Eq. (\ref{energy})  becomes:
\begin{eqnarray}
\Delta F&=&{1\over 2Z^0}{a \over 2m}\Biggl\{ \sum_{k_1,k_2}^e
	(k_1-k_2)^2 \bigl[ 
		Z^0_{Fe}n_F(k_1)n_F(k_2)\nonumber \\
&&		-{1\over Z^0_{Be}}n_B(k_1)n_B(k_2)\bigr] 
\nonumber \\ 
&&+\sum_{k_1,k_2}^o(k_1-k_2)^2\bigl[ 
		Z^0_{Fo}n_F(k_1)n_F(k_2)\nonumber \\
&&		+{1\over Z^0_{Bo}}n_B(k_1)n_B(k_2)\bigr] 
	\Biggr\} 
\end{eqnarray}
where the first sum is over even wave-vectors and the second over odd 
wave-vectors.  The resulting magnetization curve is plotted in 
Fig.\ref{MC}.  Note the smoothing of the singularity at the critical
field due to a finite $T$ and the oscillations due to a finite $L$.  We 
use the interaction parameter, $a\approx -0.34\xi$, obtained from the 
ground state numerical data so there are no free parameters in drawing 
Fig.\ref{MC}.  The agreement with the Monte Carlo data,\cite{kash} at 
the same length and temperature, L=100, T=1/100 is remarkable.

This work is partially supported by Chinese Natural Science Foundation, 
HKUGC under RGC grant HKUST6143/97P, by NSERC of Canada and by 
NSF under grant No. PHY94-07194.

\begin{references}
\bibitem{hal1} F. D. M.\ Haldane, Phys. Lett. {\bf 93A}, 464 (1983); 
		\prl {\bf 50}, 1153 (1983).
\bibitem{hus} S. R.\ White and D. A.\ Huse, \prb {\bf 48} 3844 (1993).
\bibitem{gol} O.\ Golinelli, Th.\ Jolic\oe ur and R.\ Lacaze, \prb 
		{\bf 50}, 3037(1994).
\bibitem{ken1} T. Kennedy, J. Phys: Cond. Matter {\bf 2}, 5737 (1990).
\bibitem{hagiwara} I. Hagiawara, K. Katsumata, I. Affleck,
B.I. Halperin and J.P. Renard, Phys. Rev. Lett. {\bf 65}, 3181 (1990).
\bibitem{aff1} I.\ Affleck, T.\ Kennedy, E. H.\ Lieb and H.\ Tasaki, 
		\prl {\bf 59}, 799 (1987); Commun. Math. Phys. 
		{\bf 115}, 477 (1988).
\bibitem{ng0} T. K. Ng, \prb {\bf 50}, 555(1994).
\bibitem{dit} J. F. Ditusa, S. W. Cheong, J. H. Park, G. Aeppli, 
		C. Broholm and C. T. Chen, \prl {\bf 73}, 1857(1994).
\bibitem{qin1} S. Qin, T. K. Ng, and Z. B. Su, \prb {\bf 52}, 
		12844(1995).
\bibitem{qin2} S. Qin, X. Wang and Lu Yu, \prb {\bf 56}, R14251(1997).
\bibitem{tsv} A.M. Tsvelik, Phys. Rev. {\bf B42}, 10,499 (1990).
\bibitem{aff2} I. Affleck, Phys. Rev.  {\bf B43}, 3215 (1991).
\bibitem{mag2} I. Affleck and R. A. Weston, Phys. Rev. B {\bf 45}, 
		4667(1992).
\bibitem{mag3} M. Horton and I. Affleck, cond-mat/9907431 (to appear 
		in Phys.  Rev. {\bf B}); F. Essler, cond-mat/9908186.
\bibitem{wht0} S. R.\ White, \prl {\bf 69}, 2863 (1992); \prb {\bf 48}, 
		10345 (1993); R. M. Noack and S. R. White, 
		``The density matrix renormalization group'' in 
		{\it Lecture Note in 
		Physics:} Density-Matrix Renormalization, Eds. I. 
		Peschel, X. Wang, M. Kaulke and K. Hallberg, Springer 
		(1999).
\bibitem{qin0} S. Qin, Y. L. Liu, and Lu Yu, \prb {\bf 55}, 2721(1997).
\bibitem{tak} M. Takahasi, Phys. Rev. Lett. {\bf 62}, 2313 (1990).
\bibitem{okun} K. Okunishi, Y. Hieida and Y. Akutsu,
               \prb {\bf 59}, 6806(1999).
\bibitem{kash} V. A. Kashurnikov, N. V. Prokof'ev, B. V. Svistunov, and
		M. Troyer, \prb {\bf 59}, 1162(1999).
\bibitem{sachdev} S. Sachdev, T. Senthil and R. Shankar, Phys. Rev. 
		{\bf B50}, 258 (1994).
\bibitem{aff0} see I. Affleck, in {\em Field Theory Methods and Quantum 
		Critical Phenomena}, edited by E. Brezin and J. 
		Zinn-Justin (North-Holland, Amsterdam, 1990);  I. 
		Affleck,  Nucl. Phys. B {\bf 257}, 397(1985). 
\bibitem{lieb} E.H. Lieb and W. Liniger, Phys. Rev. {\bf 130}, 
		1605 (1963); E.H. Lieb, ibid, 1616.
\bibitem{zam} A. B. Zamolodchikov and A. B. Zamolodchikov,
		Ann. Phys. {\bf 120}, 253(1979).
\bibitem{sor3} E.S. S\o rensen and I. Affleck, Phys. Rev. {\bf B51}, 
		16115 (1995).
\bibitem{sor4} E.S. S\o rensen and I. Affleck, Phys. Rev. Lett. {\bf 71}, 
		1633 (1993).
\bibitem{okun2} K. Okunishi, Phys. Rev. {\bf B60}, 4043 (1999)
\end{references}
\end{document}